# COMPLEX PATTERNS ON THE PLANE: DIFFERENT TYPES OF BASIN FRACTALIZATION IN A TWO-DIMENSIONAL MAPPING

Ricardo LÓPEZ-RUIZ [*]
Danièle FOURNIER-PRUNARET [#]

[*] Facultad de Ciencias - Edificio B,
DIIS - Área de Ciencias de la Computación,
Universidad de Zaragoza,
50009 - Zaragoza (Spain).

[#] Institut National des Sciences Appliquées,
Systèmes Dynamiques (SYD), L.E.S.I.A.,
Avenue de Rangueil, 31077 Toulouse Cedex (France).

**Abstract**

**Basins generated by a noninvertible mapping formed by two symmetrically coupled logistic maps are studied when the only parameter $\lambda$ of the system is modified. Complex patterns on the plane are visualised as a consequence of basins' bifurcations. According to the already established nomenclature in the literature, we present the relevant phenomenology organised in different scenarios:** *fractal islands disaggregation, finite disaggregation, infinitely disconnected basin, infinitely many converging sequences of lakes, countable self-similar disaggregation* **and** *sharp fractal boundary*. **By use of critical curves, we determine the influence of zones with different number of first rank preimages in the mechanisms of basin fractalization.**

## 1. INTRODUCTION

Mappings are simple models that have been extensively studied as independent objects of interest [Mira, 1987] or as ingredients of other more complex systems [Kapral, 1985; Crutchfield & Kaneko, 1987]. The unimodal one-dimensional case is well understood and the main results are summarized in [Collet & Eckmann, 1980] **and in [Mira, 1980][Mira, 1987]**. Two-dimensional endomorphisms are often obtained in different fields and, up to now, they have been mainly investigated by numerical simulations and analytical approach [Mira *et al.*, 1996].

Attractors and time evolution in two-dimensional mappings can be analysed by a set of measures such as Lyapunov exponents, power spectrum, invariant measures and fractal dimensions [Parker & Chua, 1989]. In the last years, an analytical instrument, furnished by the critical curves, has been introduced [Gumowski & Mira, 1980**; and references therein**]. This tool allows to study the bifurcations and some geometrical properties of basins, particularly the phenomenon of fractalization [Mira *et al.*, 1996**; Abraham *et al.*, 1997**].

Basins constitute an interesting object of study by themselves. If a colour is given to the basin of each attractor, we obtain a coloured **figure, which is a phase-plane visual representation of the asymptotic behaviour of the points of interest. The strong dependence on the parameters of this coloured figure generates a rich variety of complex patterns on the plane and gives rise to different types of basin fractalization as a consequence, for instance, of contact bifurcations between a critical curve segment and the basin boundary**. Taking into account the complexity of the matter and its nature, the study of these phenomena can be carried out only via the association of numerical investigations guided by fundamental considerations that can be found in [Mira *et al.*, 1996].

Different models of two-dimensional coupled maps **are scattered** in the literature of several fields (physics, engineering, biology, ecology and economics). See, for example, [Kaneko, 1983], [Yuan *et al.*, 1983], [Hogg & Huberman, 1984], [Van Biskirk & Jeffries, 1985], [Schult *et al.*, 1987], [De Sousa Vieira *et al.*, 1991] and [Aicardi & Invernizzi, 1992]. In these **works** the symmetries, dynamics, bifurcations and transition to chaos are **investigated** and interpreted within **the standards of the theory of dynamical systems.** The role played by critical curves in determining the properties and global bifurcations of basins of a double logistic map has been studied in [Gardini *et al.*, 1994].

In this paper we **explain** the basin behaviour of a two-dimensional mapping proposed initially as a coupled pair of logistic oscillators [López-Ruiz & Pérez-Garcia, 1991]. There, its dynamics, stability and attractors were presented for some range of the parameter in parallel with **the study of** two other models built under similar insights. The metric and statistical properties of that system were also computed and explained in [López-Ruiz & Pérez-Garcia, 1992]. It is our objective in the present work to continue the investigation of that system and to analyze the fractalization and parameter dependence of basins by using the technique of critical curves.

The plan is as follows. Sec. 2 is devoted to the recalls of symmetry properties and of local stability for the parameter range of interest. In Sec. 3 some notions concerning the critical curves and their application to the present case are given. In Sec. 4 we explain the different fractalization kinds, which are found in our system. Last Section contains the discussion and conclusion.

## 2. SYMMETRY AND BIFURCATIONS

A logistic map whose parameter $\mu_n$ is not fixed, $x_{n+1} = \mu_n x_n (1 - x_n)$, but itself follows a dynamics forced to remain in the interval $[1,4]$ has a **specific** dynamics [López-Ruiz, 1991]. The existence of a nontrivial fixed point at each step $n$ ensures the nontrivial evolution of the system. Thus, the different choices of $\mu_n$ give a **wide** variety of dynamical behaviours. For instance, the application of this idea produces the on-off intermittence phenomenon when $\mu_n$ is chosen random [Platt *et al.*, 1993] or the adaptation to the edge of chaos when $\mu_n$ is a constant with a small time perturbation [Melby *et al.*, 2000]. Other systems built under this mechanism are models (a), (b) and (c) presented in [López-Ruiz, 1991]. In these cases $\mu_n$ is forced to follow a logistic dynamics expanded to the whole interval $[1,4]$. The result corresponds to three different two-dimensional endomorphisms of coupled logistic maps.

We will concentrate our attention in model (b). This application can be represented by $T_\lambda : R^2 \to R^2$, $T_\lambda(x_n, y_n) = (x_{n+1}, y_{n+1})$. It takes the form:

$$x_{n+1} = \lambda(3x_n + 1)y_n(1 - y_n),$$
$$y_{n+1} = \lambda(3y_n + 1)x_n(1 - x_n),$$
(1)

where $\lambda$ is a real and adjustable parameter. In the following we shall write $T$ instead of $T_\lambda$ as the dependence on the parameter $\lambda$ is understood.

In this section, some results on the behaviour of $T$ gathered in [López-Ruiz & Pérez-Garcia, 1991] are recalled and **some** new bifurcation and dynamical aspects not collected in that publication are explained. The nomenclature is maintained.

## 2.1 *Symmetry*

This model has reflection symmetry $P$ through the diagonal $\Delta = \{(x,x), x \in R\}$. If $P(x,y) = (y,x)$ then $T$ conmutes with $P$:

$$T[P(x,y)] = P[T(x,y)].$$

Note that the diagonal is $T$-invariant, $T(\Delta) = \Delta$. In general, if $\Omega$ is an invariant set of $T$, $T(\Omega) = \Omega$, so is $P(\Omega)$ due to the conmutation property: $T[P(\Omega)] = P[T(\Omega)] = P(\Omega)$. It means that if $\{p_i, i \in N\}$ is an orbit of $T$, so is $\{P(p_i), i \in N\}$. In fact, if some bifurcation happens in the half plane below the diagonal so it is in the above half plane, and vice versa. The dynamical properties of the two halves of phase space separated by the diagonal are interconnected by the symmetry.

Also if the set $\Gamma$ verifies $P(\Gamma) = \Gamma$, so is $T(\Gamma)$. Then the $T$-iteration of a reflection symmetrical set continues to keep the reflection symmetry through the diagonal.

## 2.2 *Fixed Points, 2-Cycles and Closed Invariant Curves*

We focus our attention on bifurcations playing an important role in the dynamics, those happening in the interval $-1.545 < \lambda < 1.0843$. In this range, there exist stable attractors for each value of $\lambda$ and it has sense to study their basins of attraction.

**The restriction of $T$ to the diagonal is a one-dimensional cubic map, which is given by the equation** $x_{n+1} = \lambda(3x_n + 1)x_n(1 - x_n)$. Thus the solutions of $x_{n+1} = x_n$ are the fixed points on the diagonal,

$$p_0 = (0,0),$$

$$p_3 = \frac{1}{3}\left\{1 - \left(4 - \frac{3}{\lambda}\right)^{\frac{1}{2}}, 1 - \left(4 - \frac{3}{\lambda}\right)^{\frac{1}{2}}\right\},$$

$$p_4 = \frac{1}{3}\left\{1 + \left(4 - \frac{3}{\lambda}\right)^{\frac{1}{2}}, 1 + \left(4 - \frac{3}{\lambda}\right)^{\frac{1}{2}}\right\}$$

For $-1 < \lambda < 1$, $p_0$ is **an attractive node**. For all the rest of parameter values, $p_0$ is a **repelling node**. If $\lambda < 0$ the fixed points $p_{3,4}$ are **repelling nodes**. For $0 < \lambda < 0.75$, $p_{3,4}$ are not possible solutions. When $\lambda = 0.75$ a saddle-node bifurcation on the diagonal generates $p_{3,4}$. For $0.75 < \lambda < 0.866$, $p_3$ is a saddle point and $p_4$ is a node. In this parameter interval, **the whole diagonal segment between $p_3$ and $p_4$ is locus of points belonging to heteroclinic trajectories connecting the two fixed points.**

When $\lambda = \sqrt{3}/2 \cong 0.866$ the point $p_4$ changes its stability via a pitchfork bifurcation generating two new stable fixed points $p_{5,6}$ outside the diagonal. These points are obtained by solving the quadratic equation $\lambda(4\lambda + 3)x^2 - 4\lambda(\lambda + 1)x + 1 + \lambda = 0$. The solutions are :

$$p_5 = \left( \frac{2\lambda(\lambda+1) + \sqrt{\lambda(\lambda+1)(4\lambda^2 - 3)}}{\lambda(4\lambda + 3)}, \frac{2\lambda(\lambda+1) - \sqrt{\lambda(\lambda+1)(4\lambda^2 - 3)}}{\lambda(4\lambda + 3)} \right)$$

$$p_6 = \left( \frac{2\lambda(\lambda+1) - \sqrt{\lambda(\lambda+1)(4\lambda^2 - 3)}}{\lambda(4\lambda + 3)}, \frac{2\lambda(\lambda+1) + \sqrt{\lambda(\lambda+1)(4\lambda^2 - 3)}}{\lambda(4\lambda + 3)} \right).$$

For $\lambda = 0.957$, these two symmetric points lose stability via a Neimark-Hopf bifurcation. Each point $p_{5,6}$ gives rise a to a stable closed invariant curve. The size of these symmetric invariant curves grow when $\lambda$ increases into the interval $0.957 < \lambda < 1$, and, for some values of $\lambda$, frequency locking windows are obtained.

The **period two cycles** of the form $(x,0) \leftrightarrow (0, y)$ play an important role when $\lambda < 0$, and are found by solving the cubic equation: $\lambda^3 x^3 - 2\lambda^3 x^2 + (\lambda^3 + \lambda^2)x + 1 - \lambda^2 = 0$. The solutions are

$$p_1 = \left( \frac{\lambda - 1}{\lambda}, 0 \right) \leftrightarrow p_2 = \left( 0, \frac{\lambda - 1}{\lambda} \right),$$

$$p_7 = \left( \frac{(\lambda+1) - \sqrt{(\lambda+1)(\lambda-3)}}{2\lambda}, 0 \right) \leftrightarrow p_8 = \left( 0, \frac{(\lambda+1) + \sqrt{(\lambda+1)(\lambda-3)}}{2\lambda} \right),$$

$$p_9 = \left( \frac{(\lambda+1) + \sqrt{(\lambda+1)(\lambda-3)}}{2\lambda}, 0 \right) \leftrightarrow p_{10} = \left( 0, \frac{(\lambda+1) - \sqrt{(\lambda+1)(\lambda-3)}}{2\lambda} \right).$$

**Observe that the restriction of the map $T^2$ to the axes is invariant and reduces to the double logistic map $x_{n+1} = f^2(x_n)$ with $f(x) = \lambda x(1-x)$, so that its dynamics is completely known and it gives rise to a known cyclic behaviour on the axes for the map $T$.**

The 2-cycle $(p_1, p_2)$ exists for every parameter value and is unstable for every value of $\lambda$.

The pair of 2-cycles $(p_7, p_8)$ and $(p_9, p_{10})$ has a limited existence to the parameter intervals : $\lambda > 3$ or $\lambda < -1$. For $\lambda = 3$, they appear simultaneously by a pitchfork bifurcation from the period 2-orbit $(p_1, p_2)$. **They are always unstable for $\lambda > 3$. In the range $-1.45 < \lambda < -1$, they are stable. This pair of 2-cycles grows from the origin $p_0$ when $\lambda = -1$. It is the result of a codimension-two bifurcation giving rise also to a repelling 2-cycle in the direction on the invariant diagonal.**

## 2.3 *Transition to Chaos*

We follow the presentation of the former section. Remark that the considered system presents two different route to chaos depending on if $\lambda > 0$ or if $\lambda < 0$.

$\underline{\lambda > 0}$: For $\lambda$ slightly larger than 1, the two closed invariant curves approach the stable **invariant set** of the hyperbolic point $p_4$ on the diagonal (Fig. 1a). At first sight, for $\lambda \approx 1.029$, the system still seems quasi-periodic but a finer analysis reveals the **fingerprints** of chaotic behaviour. Effectively, the **crossing of the invariant curves with $LC_{-1}^{(2)}$ produces a folding process in the two invariant sets (Fig. 1b) cf. [Frouzakis & *al*., 1997], which gives rise to the phenomenon of weakly chaotic rings when the invariant set intersects itself (Fig. 1c) (cf. [Mira & *al*., 1996] p529). The existence of weakly chaotic ring is also confirmed by the positivity of the largest Liapunov exponent for the considered parameter values as it was computed and shown in Fig. 3(a) of [López-Ruiz & Pérez-Garcia, 1992]. For $\lambda \approx 1.032$, the tangential contact of the two symmetric invariant sets with the stable set of the saddle $p_4$ on the diagonal leads to the disappearance of those two weakly chaotic rings. Just after the contact, infinitely many repulsive cycles appear due to the creation of homoclinic points** and a single and symmetric chaotic attractor appears (Fig. 2a). For $1.032 < \lambda < 1.0843$, this chaotic invariant set folds strongly around $p_4$, and the dynamics becomes very complex (Fig. 2b). When the limit value $\lambda = 1.084322$ is reached, the chaotic **area** becomes tangent to its basin boundary, the iterates can escape to infinite and the attractor disappears by a contact bifurcation ([Mira & *al*., 1987], chap. 5) (Fig. 9d).

$\underline{\lambda < 0}$: The pair of 2-cycles $(p_7, p_8)$ and $(p_9, p_{10})$ are the germ of two period doubling cascades of bifurcations, respectively, one of them below the diagonal and the other one above it. Thus, for $\lambda = -1.45$, a pair of stable 4-cycles is created **on the axes** when the 2-cycle pairs $(p_7, p_8)$ and $(p_9, p_{10})$ loose their stability **after that an eigenvalue of each 2-cycle crosses through the value -1**. For $\lambda = -1.5$**, each 4-cycle bifurcates and a stable 8-cycle pair grow outside the axes.** For $\lambda = -1.51$, a new flip bifurcation generates the pair of 16-cycles, and so on until the dynamics **becomes chaotic**. For $\lambda = -1.5131$, the cascade of flip bifurcations has finished and a symmetric pair of chaotic **areas** has merged in the region around $p_0$ **(see Fig. 3a-b for $\lambda = -1.52$)**. When $\lambda = -1.545$ the chaotic **areas** have contact with their basin boundaries and they disappear.

## 3. STABLE ATTRACTORS

For the sake of clarity, we summarise the dynamical behaviour of the system explained above when the parameter is inside the interval $-1.545 < \lambda < 1.0843$. The different parameter regions where the mapping has stable attractors are given in two tables, one of them for $\lambda > 0$ and the other one for $\lambda < 0$.

$\underline{\lambda > 0}$:

| INTERVAL | NUMBER OF ATTRACTORS | ATTRACTORS |
|---|---|---|
| $0 < \lambda < 0.75$ | 1 | $p_0$ |
| $0.75 < \lambda < 0.866$ | 2 | $p_0$, $p_4$ |
| $0.866 < \lambda < 0.957$ | 3 | $p_0$, $p_{5,6}$ |
| $0.957 < \lambda < 1$ | 3 | $p_0$, pair of invariant **closed** |

| | | curves |
|---|---|---|
| $1 < \lambda < 1.03$ | 2 | pair of invariant **closed** curves |
| $1.03 < \lambda < 1.032$ | 2 | pair of **weakly chaotic rings** |
| $1.032 < \lambda < 1.0843$ | 1 | symmetric chaotic attractor **(or frequency lockings)** |

$\underline{\lambda < 0}$:

| INTERVAL | NUMBER OF ATTRACTORS | ATTRACTORS |
|---|---|---|
| $-1 < \lambda < 0$ | 1 | $p_0$ |
| $-1.45 < \lambda < -1$ | 2 | $(p_7, p_8), (p_9, p_{10})$ |
| $-1.5 < \lambda < -1.45$ | 2 | pair of 4-cycles |
| $-1.51 < \lambda < -1.5$ | 2 | pair of 8-cycles |
| $-1.512 < \lambda < -1.51$ | 2 | pair of 16-cycles |
| $-1.5131 < \lambda < -1.512$ | 2 | pair of $2^n$-cycles, n>5 |
| $-1.545 < \lambda < -1.5131$ | 2 | symmetric pair of chaotic areas |

## 4. CRITICAL CURVES

### 4.1 *Definitions and Properties*

An important tool used to study noninvertible maps is that of **critical curves**, introduced by Mira in 1964 (see [Mira *et al.*, 1996] for further details). The map *T* is said to be noninvertible if there exist points in state space that do not have a unique rank-one preimage under the map. Thus the state space is divided into regions, which we call $Z_i$, in which points have *i* rank-one preimages under *T*. These regions are separated by the so called critical curves *LC* , and so the number of the first rank preimages changes only when the *LC* are crossed, **except for singular points**. *LC* is the image of a set $LC_{-1}$. If the map *T* is continuous and differentiable, $LC_{-1}$ is a subset of the locus of points where the determinant of the Jacobean matrix of *T* is zero, and is the two-dimensional analogue of the set of local extrema of a one-dimensional map. When *LC* is crossed rank-one preimages appear or disappear in pairs. **(See also the glossary for different technical terms used along this work)**.

### 4.2 *Critical Curves of T*

The map *T* defined in (1) is noninvertible. It has a non-unique inverse. **Each point in the state plane can possess up to five rank-one preimages**. **The preimages can only be calculated numerically, by solving a polynomial of degree five**. The critical curve *LC* of *T* (the locus of points having at least two coincident rank-1 preimages) is the image of $LC_{-1}$, the locus of points where the Jacobean $|DT(p)|$ vanishes. That is, $LC = T(LC_{-1})$. The Jacobean matrix of $T$ depends on $(x, y)$ and the parameter $\lambda$ and has the following form :

$$DT(x, y; \lambda) = \begin{bmatrix} 3\lambda y(1-y) & \lambda(3x+1)(1-2y) \\ \lambda(3y+1)(1-2x) & 3\lambda x(1-x) \end{bmatrix}.$$

$LC_{-1}$ is the curve verifying $|DT(x, y)| = 0$. It is formed by the points $(x, y)$ that satisfy the equation :

$$27x^2y^2 + 3x^2y + 3xy^2 - 6x^2 - 6y^2 - 8xy + x + y + 1 = 0. \qquad (2)$$

$LC_{-1}$ is independent of parameter $\lambda$ and is quadratic in $x$ and $y$. Thus we can solve one of the two variables as function of the other one. For instance, the variable $y$ as function of $x$ gives us:

$$y = \frac{(-3x^2 + 8x - 1) \pm \sqrt{657x^4 - 84x^3 - 194x^2 - 4x + 25}}{54x^2 + 6x - 12}.$$

Numerical calculations allow us to discover that for every value of $x$, or equivalently for every value of $y$, there are two points belonging to two different branches of $LC_{-1}$ (Fig. 4a). Observe that $LC_{-1}$ is a quartic curve of four branches, with two horizontal and two vertical asymptotes. The branches $LC_{-1}^{(1)}$ and $LC_{-1}^{(2)}$ have as horizontal asymptote the line $y = 0.419$ and the vertical asymptote in $x = 0.419$. The other two branches, $LC_{-1}^{(3)}$ and $LC_{-1}^{(4)}$, have the horizontal asymptote in $y = -0.530$ and the vertical one is the line $x = -0.530$. The values $0.419$ and $-0.530$ are the roots of the polynomial factor, $27x^2 + 3x - 6$, that multiplies the term $y^2$ in Eq. (2). It also follows that the critical curve of rank-1, $LC^{(i)} = T(LC_{-1}^{(i)})$, $i = 1,2,3,4$, consists of four branches. The shape of $LC$ is shown in Figs. 4(b-c). $LC$ depends on $\lambda$ and separates the plane into **three** regions that are locus of points having 1, 3 or 5 distinct preimages of rank-1. **They are respectively named by $Z_i$, $i = 1,3,5$. Observe that the set of points with three preimages of rank-1, $Z_3$, is not connected and is formed by five disconnected zones in the plane (see Figs. 4b-c).** Next section is devoted to the study of this state space organisation.

Remark that $LC_{-1}$ has the reflection symmetry through the diagonal : $P(LC_{-1}) = LC_{-1}$. Then every critical curve of rank-(k+1), $LC_k = T^{k+1}(LC_{-1})$, will conserve this symmetry: $P(LC_k) = LC_k$.

### 4.3 $Z_i$ - Regions

$LC$ separates the plane into **seven** disjoint and open **zones**, which are locus of points having distinct number of preimages of rank-1 (Figs. 4b-c). We name each region depending on the number of preimages of rank-1 it has. Thus a $Z_i$-zone means the **set** of points with $i$ preimages of rank-1. Observe that **four arcs of $LC$-curve divide** the diagonal $\Delta$ in five intervals, each one **associated only with the $Z_i$-zone including it.** Then to know the number $i$ of preimages of rank-1 of each **segment** on the diagonal, **it is necessary to determine** the number of preimages of rank-1 of each $Z_i$-zone of the plane. **Then we proceed to this calculation**.

**The set of rank-1 preimages of the diagonal includes the diagonal itself and another set formed by an hyperbola of two branches. Thus, a** point on the diagonal $(x', x')$ can have preimages of rank-1 on the diagonal or on that hyperbola:

(a) <u>Preimages belonging to the diagonal</u>: It follows from Eq. (1) that the preimages $(x, x)$ of the point $(x', x')$ verify the cubic equation:

$$3\lambda x^3 - 2\lambda x^2 - \lambda x + x' = 0. \qquad (3)$$

This equation can have one or three solutions depending on $x'$ and on $\lambda$. We define the limit values: $x'_{1d} \approx 0.65\lambda$ and $x'_{2d} \approx -0.1\lambda$. If $\lambda > 0$, Eq. (3) presents one root if $x' > x'_{1d}$ or if $x' < x'_{2d}$, and three roots if $x'_{2d} < x' < x'_{1d}$. If $\lambda < 0$, Eq. (3) has one solution if $x' > x'_{2d}$ or if $x' < x'_{1d}$, and three solutions if $x'_{1d} < x' < x'_{2d}$.

(b) <u>Preimages belonging to the hyperbola</u>: Eq. (1) tells us that the image of a point $(x, y)$ is on the diagonal when $(x, y)$ verifies:

$$y = \frac{1-x}{1+3x}. \qquad (4)$$

Equation (4) represents an hyperbola of two branches with asymptotes in $x = -1/3$ and in $y = -1/3$. The point $(x', x')$ has preimages $(x, y)$ on the hyperbola if $x' = \lambda(3y+1)x(1-x)$. If we introduce relation (4) between $x$ and $y$ in the former expression we obtain the equation:

$$4\lambda x^2 + (3x' - 4\lambda)x + x' = 0. \qquad (5)$$

If the radicand of this equation is positive, the point $(x', x')$ has two preimages on the hyperbola (4), and if the radicand is negative, it has not preimages on that hyperbola. The roots of the radicand of Eq. (5) give us the behaviour of its sign:

$$9x'^2 - 40\lambda x' + 16\lambda^2 = 0. \qquad (6)$$

The roots of this equation are $x'_{1h} = 4\lambda$ and $x'_{2h} \approx 0.44\lambda$. Then, if $\lambda > 0$, Eq. (5) presents two solutions if $x' > x'_{1h}$ or $x' < x'_{2h}$, and it has no solutions when $x'_{2h} < x' < x'_{1h}$. If $\lambda < 0$, Eq. (5) has two solutions if $x' > x'_{2h}$ or if $x' < x'_{1h}$, and no solutions when $x'_{1h} < x' < x'_{2h}$.

Thus the coordinates of the points that mark the frontier between the different $Z_i$-zones on the diagonal are: $x'_{1d} \approx 0.65\lambda$ and $x'_{2d} \approx -0.1\lambda$, $x'_{1h} = 4\lambda$ and $x'_{2h} \approx 0.44\lambda$. For example, the origin $p_0$ is always in the $Z_5$-zone. It is located into the interval that determine $x'_{1d}$ and $x'_{2d}$ as extremes, then $p_0$ has three preimages on the diagonal. Also it is outside of the interval determined by the extremes $x'_{1h}$ and $x'_{2h}$, then $p_0$ has two preimages on the hyperbola (4). In fact, its preimages are $(1,1), (-1/3, -1/3)$ and $p_0$ itself on the diagonal, and $(1,0), (0,1)$ on the hyperbola. Taking and connecting the results of former paragraphs (a)-(b), the number of preimages of rank-1 of a point $(x', x')$ on the diagonal can be summarised in the following tables.

<u>$\lambda > 0$</u>:

| INTERVAL | $x' < x'_{2d}$ | $x'_{2d} < x' < x'_{2h}$ | $x'_{2h} < x' < x'_{1d}$ | $x'_{1d} < x' < x'_{1h}$ | $x' > x'_{1h}$ |
|---|---|---|---|---|---|
| NUMBER OF PREIMAGES | 3 | 5 | 3 | 1 | 3 |

$\lambda < 0$:

| INTERVAL | $x' < x'_{1h}$ | $x'_{1h} < x' < x'_{1d}$ | $x'_{1d} < x' < x'_{2h}$ | $x'_{2h} < x' < x'_{2d}$ | $x' > x'_{2d}$ |
|---|---|---|---|---|---|
| NUMBER OF PREIMAGES | 3 | 1 | 3 | 5 | 3 |

**To summarise, the many different zones $Z_i$ separated by branches of the critical curve $LC$ can be classified, according to the nomenclature established in Mira *et al.* [1996], in two different schemes. For $\lambda > 0$, the map (1) is of the type $Z_3 - Z_5 \succ Z_3 - Z_1 \prec Z_3$ and for $\lambda < 0$, the map is of the type $Z_3 \succ Z_1 - Z_3 \prec Z_5 - Z_3$.**

Observe in Figs. 4(b-c) that the branches of the critical curve $LC$ are not smooth. They exhibit two cusp points that coincide with $x'_{1h}$ and $x'_{2h}$. A cusp point can be considered as a singular point of $LC$ where three first rank preimages coincide. These points give information on the sheet structure of the map foliation. In the case of the cusp point $x'_{1h}$ the two preimages on the hyperbola (4) and the preimage on the diagonal are mixed up on the point $(-1,-1)$. Equivalently, $x'_{2h}$ has two preimages, $(0.91, 0.91)$ and $(-0.52, -0.52)$ on the diagonal and the other three preimages coincide with the point $(1/3, 1/3)$. For the map $T$ there is no bifurcation involving the cusp points and associated with the qualitative change of the sheet organisation. The only bifurcation occurs in $\lambda = 0$, and gives rise to an inversion in the plane of the sheet structure (Figs. 4b-c).

## 5. BASIN FRACTALIZATION

### *5.1 General Properties*

**The set $D$ of initial conditions that converge towards an attractor at finite distance when the number of iterations of $T$ tends toward infinity is the basin of the attracting set at finite distance. When only one attractor exists at finite distance, $D$ is the basin of this attractor.** When several attractors at finite distance exist, $D$ is the union of the basins of each attractor. The set $D$ is invariant under backward iteration $T^{-1}$ but not necessarily invariant by $T$: $T^{-1}(D) = D$ and $T(D) \subseteq D$. A basin may be connected or non-connected. A connected basin may be simply connected or multiply connected, which means connected with holes. A non-connected basin consists of a finite or infinite number of connected components, which may be simply or multiply connected [Mira *& al.*, 1994].

The closure of $D$ includes also the points of the boundary $\partial D$, whose sequences of images are also bounded and lay on the boundary itself. If we consider the points at infinite distance as an attractor, its basin $D_\infty$ is the complement of the closure of $D$. When $D$ is multiply connected, $D_\infty$ is non-connected, the holes (called lakes) of $D$ being the non-connected parts (islands) of $D_\infty$. Inversely, non-connected parts (islands) of $D$ are holes of $D_\infty$ [Mira *et al.*, 1996].

In Sec. 3, we explained that the map (1) may possess one, two or three attractors at a finite distance. The points at infinity constitute the fourth attractor of $T$. Thus, if a different colour for each different basin is given we obtain a coloured pattern with a maximum of four colours. In the present case, the phenomena of basin boundary fractalization have their origin in the competition between the attractor at infinity (**whose basin is** $D_\infty$) and the attractors at finite distance (**whose basin is** $D$). The competition between the attractors at finite distance can also fractalize the set $D$ (see Sec. 5.3.2). When a bifurcation of $D$ takes place, some important changes **appear** in the coloured **figure**, and, although the dynamical causes can not be clear, the coloured **pattern** becomes an important visual tool to analyze the behaviour of basins.

We study in detail, and with the help of graphical representations, the mechanisms of basin fractalization **for the map** $T$. The changes of the shape and evolution of $D$ are given in function of the sign of the parameter $\lambda$. Two scenarios are **distinguished**: *Scenario I* for $\lambda > 0$ and *Scenario II* for $\lambda < 0$.

## 5.2  *Scenario I*, $\lambda > 0$

When $0 < \lambda < 1.084$, we observe important modifications **of the area of** $D$**, the basin of the attracting set at finite distance. Its size decreases when** $\lambda$ **increases**. Two main contact bifurcation phenomena giving rise to a fractal basin boundary take place into this interval: fractal islands disaggregation if $0.39 < \lambda < 0.61$ and **infinitely many converging sequences of lakes** if $1.08 < \lambda < 1.084$. We proceed to explain the role played by critical curves in the bifurcations giving rise to these phenomena.

### 5.2.1  *Fractal Islands Disaggregation,* $0.39 < \lambda < 0.61$

When $\lambda$ increases from $\lambda \approx 0.39$ to $\lambda \approx 0.61$, $D$ **undergoes two bifurcations connected - non-connected and non-connected – connected. When** $D$ **is non-connected, it is made up of the immediate basin** $D_0$ **containing the single attractor** $(p_0)$ **and infinitely many small regions without connection (islands). This disaggregation is the result of infinitely many contact bifurcations, which are explained in the next paragraph. Such phenomenon can be also found in some quadratic** $Z_0 - Z_2$ **maps and has been explained in [Mira & Rauzy, 1995]. We explain for decreasing** $\lambda$ **the corresponding mechanisms (Fig. 6a-h).**

When $\lambda \approx 0.61$ (Figs. 6e-h), a first rank island is created due to the intersection of $LC^{(1)}$ with the two symmetric narrow "tongues" of $D$ located **in the third quadrant, and having the lines** $x = -1/3$ **and** $y = -1/3$ **as asymptotes.** Those points of $D$ crossing $LC^{(1)}$ from the below $Z_3$-zone to the $Z_5$-zone acquire two new preimages of rank-1. These preimages appear as shaped islands intersecting $LC_{-1}^{(1)}$ on the middle $Z_3$-zone. **Higher rank preimages of these seminal islands give rise to new smaller islands. The same mechanism of creation of new islands comes from the intersection of the existing islands with** $LC$ **arcs (Figs. 6c-d).** Thus islands crossing $LC^{(3)}$ from the middle $Z_3$-zone to the $Z_5$-zone undergo a contact bifurcation that creates a new pair of rank-1 islands intersecting $LC_{-1}^{(3)}$.

Then, (Figs. 6a-b) islands evolving in the plane from the middle $Z_3$-zone to the $Z_1$-zone across $LC^{(2)}$ give place to the aggregation of a pair of rank-1 islands located to both sides of $LC_{-1}^{(2)}$. This cascade of bifurcations, considered when $\lambda$ increases, generates a fractal pattern of non-connected islands in the vicinity of the immediate basin $D_0$ and at the both sides of its three like-axes of symmetry: $LC_{-1}^{(1)}$, $LC_{-1}^{(2)}$ and $LC_{-1}^{(3)}$ (Fig. 6c). When $\lambda \approx 0.396$, the two headlands of $D_\infty$ enclosed between $LC^{(1)}$ and the two former "tongues" disappear and an infinitely complex aggregation happens: the frontier of every island contact with the boundary $\partial D_0$ simultaneously and a connected basin $D$ is obtained finally (see contact at point A in Fig. 6a).

**Observe that $D$ has now a non-smooth boundary (Figs. 5a-b). The stable set of the period two saddle $(p_1, p_2)$ and its preimages form part of it. To find the analytical expression of the boundary is a difficult task but in this particular case, we can obtain by direct visual inspection of the figures some partial and approximated information. The coordinates axes are $T^2$-invariant: $T$ reduces on the axes to the double logistic map $x_{n+1} = f^2(x_n)$ with $f(x) = \lambda x(1-x)$, so that it gives rise to a known dynamics on the axes for the map $T$. Two apparently unbounded "tongues" escaping toward infinity are in the asymptotical direction of the lines $x = -1/3$ and $y = -1/3$ as it can be inspected in Fig. 5a. These two lines are the rank-1 preimages of the coordinates axes, which are also asymptotes of other two similar "tongues". Then an analytical approximation of the frontier of the former "tongues" can be obtained by backward iteration of the boundary of the latter "tongues". After some straightforward calculation, it can be found that the map $T^2$ generates stable dynamics on the axes when the initial conditions $(x_0, 0)$ or $(0, y_0)$ are lying on the intervals, $1 - 1/\lambda < x_0 < 1/\lambda$ or $1 - 1/\lambda < y_0 < 1/\lambda$, for $\lambda < 1$. We call these intervals $\delta$ and $\delta'$, respectively. Preimages of different ranks higher than 1 of points in the vicinity of $\delta$ and $\delta'$ generate a set of curves that come close to and that can give us a rough idea of the fractal structure of the basin boundary at the "tongues". For instance, the piece of the curve, $x = -1/3 + 1/(9\lambda y^2)$ with $0 < y < \lambda^{-3}$, close to the asymptote $x = -1/3$, is a preimage of rank-2 of points in the vicinity of the positive part of $\delta'$ (Fig. 5b).**

### 5.2.2 *Finite Disaggregation and Infinitely Disconnected Basin*, $0.7 < \lambda < 1.032$

**When $0.7 < \lambda < 1$, $D$ seems to be formed by the square $D_0 \equiv [1,0] \times [0,1]$, which contains the attracting set at finite distance, and four small like-triangled regions linked to the square by four narrow arms (Fig. 7a). These arms shrink when $\lambda$ approaches 1, and disappear for $\lambda = 1$ when the origin $p_0$ undergoes a transcritical bifurcation. The main part of $D$ is then a disconnected pattern of five components: the square $D_0$, a triangle-shaped component located in a $Z_3$ neighbourhood of the vertex point $(-1/3, -1/3)$ (preimage of rank-1 of the point $p_0$), and the three triangle-shaped regions that are preimages of rank-1 of the latter component (Fig. 7b). As there is no region $Z_0$, an infinite sequence of rank-1 preimages exists and gives rise in this case to the appearance of much smaller islands of preimages. They can be observed by enlargement of the figures (see some of them in Fig. 7c).**

**Points $(1,0)$ and $(0,1)$ cross through $LC^{(2)}$ when $\lambda = 1$. When $\lambda > 1$, it makes appear two regions $S_1^1$ and $S_2^1$, which are part of $D_\infty$ and are located in a $Z_3$ zone. The rank-one preimages of $S_1^1$ and $S_2^1$, respectively $S_1^{-1}$ and $S_2^{-1}$ are two new semicircular regions and intersect $LC_{-1}^{(2)}$. They are located in the vicinity of points $(1, 0.5)$ and $(0.5, 1)$, preimages of $(1,0)$ and $(0,1)$ (Fig. 7d).**

When $0.866 < \lambda < 1.032$, $T$ possesses two or three attractors (see Sec. 3). If the basin of each attractor is coloured in a different way, we observe that each basin included in $D_0$ is non-connected and is made up of infinitely many components. The boundary of each connected component belongs to the stable set of unstable points [Gardini *et al.*, 1994]. Figs. 7c & 8 show $D_0$ for $\lambda = 1$. In this case, two closed invariant curves attract the dynamics and two different colours define the infinitely disconnected basin pattern.

### 5.2.3 *Infinitely Many Converging Sequences of Lakes,* $1.08 < \lambda < 1.084$

When $\lambda$ increases, the two semicircular shaped zones of $D_\infty$, $S_1^{-1}$ and $S_2^{-1}$, located in the immediate basin $D_0 \equiv [1,0] \times [0,1]$ in the neighbourhood of points $(1, 0.5)$ and $(0.5, 1)$, grow in size. For $\lambda \approx 1.0801$, the basin undergoes a contact bifurcation. $D_\infty$ crosses through $LC^{(2)}$ and two bays (headlands of $D_\infty$), $H_{01}$ and $H_{02}$, are created. Their rank-1 preimages, $H_{01}^{(1)}$ and $H_{02}^{(1)}$, are holes (lakes) intersecting $LC_{-1}^{(2)}$ into the middle $Z_3$-region (Fig. 9a). Rank-1 preimages of the latter holes generate four new lakes in $D_0$, $H_{0i}^{(21)}$ and $H_{0i}^{(22)}$, $i = 1,2$ (Fig. 9b). Preimages with increasing rank give rise to an arborescent sequence of lakes. Inside $D_0$, the accumulation points of this infinite sequence of holes are the two unstable foci $p_{5,6}$ and their rank-1 preimages inside the immediate basin. Outside $D_0$, the other two rank-1 preimages of $p_{5,6}$ are limit points of the arborescent sequence of holes generated on the bigger triangle island with vertex point $(-1/3, -1/3)$ (Fig. 9c).

When $\lambda \approx 1.0806$, $H_{01}^{(21)}$ and $H_{02}^{(21)}$ cross through $LC^{(2)}$ (Fig 9b). This new contact bifurcation is the germ of a new arborescent and spiralling sequence of lakes converging towards the same accumulation points. When $\lambda$ increases values, new holes intersect $LC^{(2)}$ and give rise to new holes crossing through $LC_{-1}^{(2)}$ and new sequences of lakes converging towards the unstable foci $p_{5,6}$ and their preimages. Due to the fact that the preimages have a finite number of accumulation points, the structure is not fractal (Fig 9b). A similar phenomenology has been found and studied in $Z_0 - Z_2$ maps [Mira *et al.*, 1994].

When $\lambda$ increases ($\lambda \approx 1.0835$), the chaotic attractor, which is limited by arcs of $LC_n$ curves, is destroyed by a contact bifurcation with its basin boundary. A new dynamical state arises. The infinite number of unstable cycles and their rank-n images belonging to the existing chaotic area before the bifurcation define a strange repulsor which manifests itself by chaotic transients (Fig. 9e). For $\lambda \approx 1.085$ the basin pattern disappears definitively.

### 5.3 *Scenario II,* $\lambda < 0$

When $-1.545 < \lambda < 0$, **the size of the domain of bounded iterated sequences decreases** when $\lambda$ decreases. Two different qualitative phenomena arise in this case for decreasing $\lambda$: **countable self-similar disaggregation** if $-1.2 < \lambda < -0.85$ **and sharp boundary fractalization** if $-1.545 < \lambda < -1.45$. We analyze the patterns generated in next paragraphs.

## 5.3.1 Countable Self-similar Disaggregation, $-1.2 < \lambda < -0.85$

In the interval $-1 < \lambda < 0$, the origin $p_0$ is the only attractor. The basin $D$ is connected. It presents several **"tongues"**: unbounded segments escaping apparently towards infinity and having the vertical lines $x = -1/3$ and $x = 1$, and the horizontal **straight lines** $y = -1/3$ and $y = 1$ as asymptotes (see Sec. 5.2.1). When $\lambda \approx -0.85$, $LC^{(2)}$ crosses through the boundary $\partial D$, creating a small **headland** $H_0$ of $D_\infty$ trapped between $LC^{(2)}$ itself and $\partial D$ (Figs. 10a-b-c). A rank-1 preimage of this small region creates a big **lake** $H_0^{-1}$ in $D$ intersecting $LC_{-1}^{(2)}$. **Other lakes of smaller size appear on the tongues as preimages of** $H_0^{-1}$. The sequence of rank-n preimages of these holes generates an infinite sequence of lakes along the diagonal and on the tongues (Figs. 10c). **The basin becomes multiply connected.**

**New headlands of $D_\infty$, $H_1$ and $H_2$, are created between $LC^{(2)}$ and $\partial D$ when $\lambda$ decreases. It gives rise to new arborescent sequences of holes located close to the former ones (Figs. 10d-e). Afterwards, the aggregation of neighbour headlands of $D_\infty$ gives rise to the aggregation of their images lakes in the interior of $D$. Other bays open in the sea, due to the crossing of $LC^{(2)}$ through $\partial D$, giving rise to the transformation of their images lakes in roadsteads (Figs. 10f-g-h). For $\lambda \approx -1.05$, the combination of these two phenomena breaks $D$ in a set of infinitely many components located on the diagonal and on the tongues, as a consequence of that $LC^{(2)}$ is totally contained in $D_\infty$ and the former headlands of $D_\infty$ are open in the sea. The new basin is a set with a countable number of self-similar parts. If we make enlargments of some parts, they seem identical to the former one in the sequence (Figs. 10i-j-k). It means that the basin $D$ can be seen as a collection of arbitrarily small pieces, each of which is an exact scaled down version of the entire basin.**

If $l_n$ is the size of component $n$ of this sequence, the scaling factor $r$ defined as $r \approx \lim_{n \to \infty} (l_n / l_{n+1})$ takes the value $r \approx 7.43$.

To calculate this scaling factor we remark that the sequence of diagonal preimages $x_n$ (with $x_1 = 1$ and $x_n > 1$ for all $n > 1$) of the origin $p_0 (x_0 = 0)$ are located one by one in the core of each component of the self-similar basin. For $\lambda = -1$, this sequence $\{x_n\}$ can be calculated with the restriction of recurrence (1) on the diagonal:

$$x_n = 3x_{n+1}^3 - 2x_{n+1}^2 - x_{n+1}, \qquad (7)$$

where $x_0 = 0$, $x_1 = 1, \ldots$. The accumulation point $x_\infty$, where the sequence $\{x_n\}$ converges, is obtained as the fixed point of the former recurrence:

$$x_n = x_{n+1} \to x_\infty = \frac{1 + \sqrt{7}}{3} \approx 1.21.$$

**Observe that $x_\infty$ is the coordinate of the fixed point $p_4$ on the diagonal. That is, the sequence $\{x_n\}$ is an heteroclinic orbit between $p_0$ and $p_4$. The difference between two consecutive terms, $x_n - x_{n-1}$, gives us, approximately, the size $l_n$ of the n-th component of the self-similar** basin. The scaling factor $r$ is obtained linearizing Eq. (7) around the fixed point $x_\infty$. The result is:

$$r \approx \lim_{n \to \infty} \frac{l_n}{l_{n+1}} = \lim_{n \to \infty} \frac{x_n - x_{n-1}}{x_{n+1} - x_n} = 9x_\infty^2 - 4x_\infty - 1 \approx 7.43$$

Observe that $r$ is the diagonal multiplier of the unstable fixed point $p_4$.

### 5.3.2 Sharp Fractal Boundary, $-1.545 < \lambda < -1.45$

Two stable period 2 cycles are created for $\lambda = -1$, as consequence of a codimension-two bifurcation in $p_0$ (see Sec. 3). These 2-cycles persist stable in the parameter range $-1.45 < \lambda < -1$. Afterwards the system undergoes a double cascade of flip bifurcations that gives rise to a symmetric pair of chaotic **areas** for $\lambda \approx -1.5131$ (see Figs. 3). Then two colours define the basin of attractors at finite distance, $D = D_1 \cup D_2$, in the interval $-1.545 < \lambda < -1$, being $D_1$ and $D_2$ the basin of each attractor (Fig. 10f & following).

For $\lambda \approx -1.45$, the critical curve $LC^{(3)}$ intersects the boundary separating $D_1$ and $D_2$, creating headlands of one basin between $LC^{(3)}$ itself and the boundary of the other basin (Fig. 11a-b). After this contact bifurcation a cascade of islands of basin $D_1$ into $D_2$, and vice versa, is generated. A fractal pattern is created. Observe that the fractalization is located in very determined zones of $D$. Each of these regions is limited apparently by two smooth curves (whose equations are unknown for us) intersecting on the diagonal. The origin and its preimages are on the frontier of these regions as accumulation points (Fig. 11a-b). Although islands of $D_1$ are spread out over $D_2$, and vice versa, in the fractal zones, these are not riddled basins in the sense defined in [Alexander *et al.*, 1992]. We can always find on those regions a disk, which intersects one of the basins in a set of positive measure but does not intersect the other basin.

A similar behaviour is repeated for $\lambda \approx -1.485$ but with the basin of infinity. $LC^{(3)}$ crosses through the boundary of this basin and a cascade of islands of $D_\infty$ is created in the regions where $D_1$ and $D_2$ are intermingled (Fig. 11c-d). Observe that now the origin and the sequence of its preimages are cusp points located on the frontier of $D$. For $\lambda$ decreasing, many different $D_\infty$-islands have contacts with the external boundary of $D$ and they open in the sea (Fig. 11e). This arborescent sequence of roadsteads gives place to a pattern with a sharp fractal boundary.

As in Sec. 5.2.3, for $\lambda \approx -1.545$, a contact between $\partial D$ and the frontier of the chaotic area destroy the attractor and a strange repulsor takes its place. The dynamics derives towards infinity giving rise to the disappearance of the basin.

## 6. CONCLUSION

Basins of a noninvertible two-dimensional mapping formed by two symmetrically coupled logistic maps have been analysed. Two approaches have been used to study the bifurcations leading to their fractalization. On one hand, the geometrical and coloured representations of basins as a phase-plane visual instrument allowing us a qualitative focus, and, by the other hand, a more precise and analytical approach furnished by the tool of critical curves.

The map is of degree five and a region with five first rank preimages exists for every value of the parameter $\lambda$. Also there are regions with one and three rank-1 preimages. The many different zones $Z_i$ separated by branches of the critical curve $LC$ can be classified, according to the nomenclature established in [Mira *et al.*, 1996], in $Z_3 - Z_5 \succ Z_3 - Z_1 \prec Z_3$ type for $\lambda > 0$ and, $Z_3 \succ Z_1 - Z_3 \prec Z_5 - Z_3$ for $\lambda < 0$. The symbols ">" and "<" indicate the existence of two cusp points located on the frontiers separating the $Z_1$ and $Z_5$ zones from two of the $Z_3$ zones. We point out the non-existence of a zone with zero preimages, then every point on the plane has an infinite sequence of preimages.

Although the system has only one parameter, it gives rise to a rich basin behaviour, which generates many different complex patterns on the plane. Considering the complexity of all this phenomenology and being aware of the difficulty of such an attempt, we have organised every different type of basin fractalization present in this system in the following way: *fractal islands disaggregation* for $0.39 < \lambda < 0.61$, *finite disaggregation* and *infinitely disconnected basin* for $0.7 < \lambda < 1.032$, *infinitely many converging sequences of lakes* for $1.08 < \lambda < 1.084$, *countable self-similar disaggregation* for $-1.2 < \lambda < -0.85$ and *sharp fractal boundary* for $-1.545 < \lambda < -1.45$. Each of these fractalization types have common features with those present in the simplest and well-studied case of $Z_0 - Z_2$ maps. We want to call the attention on the set of patterns found in the interval $-1.2 < \lambda < -0.85$. It constitutes a new and easily calculable example of a self-similar countable set on the plane. It can be seen as a collection of arbitrarily small pieces, each of which is an exact scaled down version of the entire basin. The scaling factor has been calculated by analytical direct inspection of recurrence equations.

Finally we remark that, in general, basin fractalization does not imply the existence of a chaotic attractor in the system. In fact, two of the fractalization types analysed in our mapping present a periodic underlying dynamical behaviour.

**Acknowledgements**

We would like to thank Prof. Mira and Dr. Taha for their helpful comments. We also acknowledge the significantly improvement of this article after carefully including all the referee's remarks. One of the authors (R. L.-R.) wishes to thank the group of Systèmes Dynamiques at INSA (Toulouse), where the most part of this work was done, for his kind hospitality , and the Program Europa of CAI-CONSI+D (Zaragoza) for financial support.

--------------------------

# GLOSSARY

**INVARIANT:** A subset of the plane is invariant under the iteration of a map if this subset is mapped exactly onto itself.

**ATTRACTING:** An invariant subset of the plane is attracting if it has a neighbourhood every point of which tends asymptotically to that subset or arrives there in a finite number of iterations.

**CHAOTIC AREA:** An invariant subset that exhibits chaotic dynamics. A typical trajectory fills this area densely.

**CHAOTIC ATTRACTOR:** A chaotic area, which is attracting.

**BASIN:** The basin of attraction of an attracting set is the set of all points, which converge towards the attracting set.

**IMMEDIATE BASIN:** The largest connected part of a basin containing the attracting set.

**ISLAND:** Non-connected region of a basin, which does not contain the attracting set.

**LAKE:** Hole of a multiply connected basin. Such a hole can be an island of the basin of another attracting set.

**HEADLAND:** Connected component of a basin bounded by a segment of a critical curve and a segment of the immediate basin boundary of another attracting set, the preimages of which are islands.

**BAY:** Region bounded by a segment of a critical curve and a segment of the basin boundary, the successive images of which generate holes in this basin, which becomes multiply connected.

**COAST:** Basin boundary.

**SEA:** An open domain of divergent iterated sequences.

**ROADSTEAD:** The coast situation obtained after a bifurcation opening a lake in a sea.

**DEGREE OF A NONINVERTIBLE MAP:** Maximum of rank-one preimages generated by the map.

**CONTACT BIFURCATION:** Bifurcation involving the contact between the boundaries of different regions. For instance, the contact between the boundary of a chaotic attractor and the boundary of its basin of attraction or the contact between a basin boundary and a critical curve *LC*.

**AGGREGATION:** The situation obtained after that two or more disconnected components of a basin form a single connected component.

**EXTERNAL BOUNDARY:** Boundary of the immediate basin (and other basin components if they exist) obtained by taking out the lakes.

**CUSP POINT:** (1) Point on the critical curve *LC* where three first rank preimages coincide. (2) Repelling node belonging to the external basin boundary, the eigenvalues, $\eta_1, \eta_2$, of which satisfy $\eta_1 > 1$, $\eta_2 < -1$ **and** $|\eta_2| > |\eta_1|$.

**SHARP FRACTAL BOUNDARY:** Basin boundary, the external boundary of which is made up of arcs having a fractal structure containing an arborescent sequence of preimages of a cusp point. Thus infinitely many cusp points belong to the boundary.

# Figure Captions

**Fig. 1:** **(a)** Attractive closed invariant curves for $\lambda = 1.03$, and **(b)** Enlargment of (a), **(c)** Weakly chaotic rings limited by segments of critical curves $LC_n$.

**Fig. 2:** **(a)** Symmetric chaotic attractor for $\lambda = 1.083$. **(b)** Complex folding process around $p_4$ for $\lambda = 1.08$.

**Fig. 3:** **(a)** Pair of chaotic areas for $\lambda = -1.52$ as a result of two period doubling cascades. **(b)** Enlargment of a chaotic area.

**Fig. 4:** **(a)** Critical curves $LC_{-1}^{(i)}$, $i = 1,2,3,4$. **(b)-(c)** Critical curves $LC^{(i)}$, $i = 1,2,3,4$, for $\lambda = 0.3$ and for $\lambda = -0.3$, respectively. Observe the different $Z_j$-zones, $j = 1,3,5$.

**Fig. 5:** **(a)** Basin $D$ for $\lambda = 0.3$. Observe the "tongues" having as asymptotes the lines: $x = -1/3$, $y = -1/3$, $x = 0$ and $y = 0$. **(b)** Detail of the fractal structure of the tongues.

**Fig. 6:** **(a)** Collective islands disaggregation of basin $D$ for $\lambda = 0.396$. **(b)** Observe the fractal pattern of islands to both sides of the three like-axes of symmetry: $LC_{-1}^{(1)}$, $LC_{-1}^{(2)}$ and $LC_{-1}^{(3)}$ for $\lambda = 0.45$. **(c)-(d)-(e)-(f)** First rank islands generated from the intersection of $LC^{(1)}$ with basin "tongues", for $\lambda = 0.6$, **(g)-(h)** $\lambda = 0.7$

**Fig. 7:** **(a)** Basin $D$ formed by the square $D_0 \equiv [1,0] \times [0,1]$, which contains the attractors, and four small like-triangle regions linked to the square by four narrow arms for $\lambda = 0.9$. **(b)** $D$ has five disconnected components for $\lambda = 1$. **(c)** Enlargment of (b), **(d)** Observe the two semicircular shaped regions of $D_\infty$, $S_1^{-1}$ and $S_2^{-1}$, intersecting $D_0$ for $\lambda = 1.03$.

**Fig. 8:** $D_0$ for $\lambda = 1$. The two colours correspond to the non-connected basins of two attractive closed invariant curves.

**Fig. 9:** **(a)** $D_0$ for $\lambda = 1.0803$: first rank holes $H_{01}^{(1)}$ and $H_{02}^{(1)}$ (and higher rank preimages holes) of the bays $H_{01}$ and $H_{02}$, respectively. **(b)-(c)** New arborescent sequence of holes created from the crossing of $H_{01}^{(21)}$ and $H_{02}^{(21)}$ with $LC^{(2)}$, **(d)** Chaotic attractor and its basin **(e)** Strange repulsor for $\lambda = 1.084$.

**Fig. 10:** Countable self-similar disaggregation of basin $D$ (explanation in the text): **(a)** $\lambda = -0.85$, **(a)** $\lambda = -0.85$ (enlargement), **(c)** $\lambda = -0.93$, **(d)** $\lambda = -0.97$, **(e)** $\lambda = -0.97$ (enlargement), **(f)** $\lambda = -1.01$, and, **(g)-(h)** $\lambda = -1.01$ (enlargement), **(i)** $\lambda = -1.4$, **(j)-(k)** $\lambda = -1.4$ (enlargement).

**Fig. 11:** Sharp boundary fractal of basin $D$ (explanation in the text): **(a)** $\lambda = -1.46$, **(b)** $\lambda = -1.46$ (enlargement), **(c)** $\lambda = -1.5$, **(d)** $\lambda = -1.5$ (enlargement), and, **(e)** $\lambda = -1.54$, **(f)** $\lambda = -1.54$ (enlargement).